\documentclass[lettersize,journal]{IEEEtran}
\usepackage{amsmath,amsfonts}
\usepackage{algorithmic}
\usepackage{subfigure}
\usepackage{algorithm}
\usepackage{array}
\usepackage{textcomp}
\usepackage{stfloats}
\usepackage{url}
\usepackage{verbatim}
\usepackage{multirow}
\usepackage{booktabs}
\usepackage{graphicx}
\usepackage[table,x11names]{xcolor}
\usepackage{cite}
\usepackage{caption}
\usepackage{subcaption}
\usepackage{svg}
\usepackage{pgfplots}
\usepackage{tikz}
\usepackage{circuitikz}
\usepackage{hyperref}
\usepackage{xspace}
\usepackage{scalerel}
\usepackage{tikz}
\usetikzlibrary{svg.path}

\definecolor{orcidlogocol}{HTML}{A6CE39}
\tikzset{
  orcidlogo/.pic={
    \fill[orcidlogocol] svg{M256,128c0,70.7-57.3,128-128,128C57.3,256,0,198.7,0,128C0,57.3,57.3,0,128,0C198.7,0,256,57.3,256,128z};
    \fill[white] svg{M86.3,186.2H70.9V79.1h15.4v48.4V186.2z}
                 svg{M108.9,79.1h41.6c39.6,0,57,28.3,57,53.6c0,27.5-21.5,53.6-56.8,53.6h-41.8V79.1z M124.3,172.4h24.5c34.9,0,42.9-26.5,42.9-39.7c0-21.5-13.7-39.7-43.7-39.7h-23.7V172.4z}
                 svg{M88.7,56.8c0,5.5-4.5,10.1-10.1,10.1c-5.6,0-10.1-4.6-10.1-10.1c0-5.6,4.5-10.1,10.1-10.1C84.2,46.7,88.7,51.3,88.7,56.8z};
  }
}

\newcommand\orcidicon[1]{\href{https://orcid.org/#1}{\mbox{\scalerel*{
\begin{tikzpicture}[yscale=-1,transform shape]
\pic{orcidlogo};
\end{tikzpicture}
}{|}}}}

\begin{document}
\title{SRAM Based Digital Custom Compute Engine for Improved Area Efficiency of AI Hardware}

\author{Narendra Singh Dhakad \orcidicon{0000-0003-2848-1785} and Santosh Kumar Vishvakarma \orcidicon{0000-0003-4223-0077}
\thanks{Narendra Singh Dhakad and Santosh Kumar Vishvakarma are associated with Nanoscale Devices, VLSI Circuit and System Design (NSDCS) Lab, Department of Electrical Engineering, Indian Institute of Technology Indore, MP 453552, India.}
}

\maketitle
\begin{abstract}
This paper presents a novel architecture utilizing a 10T SRAM cell for XNOR-based in-memory computing, aimed at mitigating the extensive routing challenges typically encountered in conventional in-memory computing systems. By integrating a full adder between in-memory multiplication cells, the proposed design achieves a 50\% reduction in routing complexity. The architecture performs multiply-accumulate (MAC) operations using XNOR computation optimized for binary neural networks (BNNs). Additionally, a 14T-based full adder is employed to construct an N-bit ripple carry adder in the adder tree, significantly reducing the area compared to traditional 28T-based CMOS designs. The 10T SRAM XNOR computation further enhances the latency for MAC operations. The proposed approach reduces the latency and area overhead, improving the overall hardware's area efficiency by 2.67$\times$ compared to the state-of-the-art. 
\end{abstract}


\section{Introduction}
\begin{figure}[t]
    \centering
    \includegraphics[width=\linewidth]{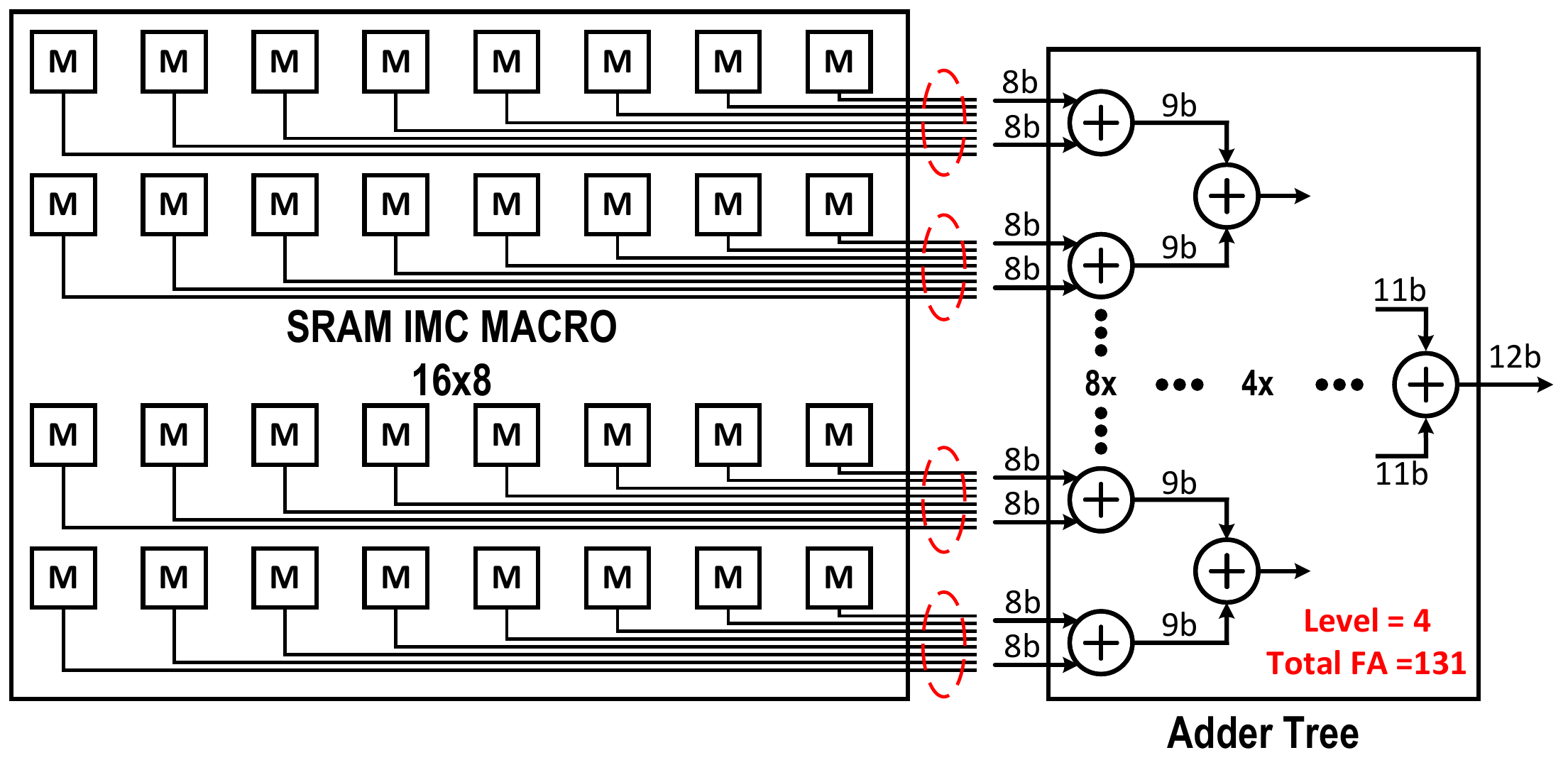}
    \caption{Conventional digital IMC architecture. Here, the multiplication of weight and bias is performed inside memory, and accumulation is done through an adder tree. }
    \label{Conv_bittree}
\end{figure}

Over the last few years, neural network algorithms have been explored for various AI tasks. With the advancement of AI, workloads, and data, the efficient architecture to handle the computation tasks is needed for the hours. Many hardware accelerators have been proposed in the past for efficient computations. General-purpose hardware accelerators perform multiply and accumulate (MAC) locally implemented using digital designs \cite{neuro}. However, the digital approach consumes more area and power. Also, to perform convolution, a huge amount of data need to be fetched multiple times, which consumes huge energy and increases the latency. Fig. \ref{Conv_bittree} shows the general-purpose digital compute engine, which uses memory to store the weights, a controller circuit to handle memory and compute controls, and a compute engine. The compute engine has one multiplier, adder, and register and performs MAC (Multiply and Accumulate) operations. In this approach, multiple input and weight channels are accumulated through a partial sum approach, where the output compute data for one cycle is stored in the register and added to the next cycle's compute data. However, designing the compute engine digitally consumes a lot of area, energy, and power, in addition to the huge data transfer to and from memory.

% \begin{figure}[t]
%     \centering
%     \includegraphics[width=\linewidth]{Figure/LCE.pdf}
%     \caption{General purpose digital compute engine}
%     \label{digitl_compute}
% \end{figure}

% \begin{figure}[t]
%     \centering
%     \includegraphics[width=\linewidth]{Figure/IMCU.pdf}
%     \caption{General purpose in-memory compute engine}
%     \label{IMCompute}
% \end{figure}

To mitigate the issues in the digital design approach, in-memory computing (IMC) has emerged as a prominent solution to the data transfer issue in the past few years. MAC can be performed within the memory unit \cite{ISSCC2023} \cite{Das}. IMC provides better energy efficiency and reduces the latency of the computations. Fig. \ref{Conv_bittree} shows the in-memory compute engine (IMCU) based approach where the computed output is accumulated through an adder tree. The IMC solutions available in the market provide good compute efficiency but face the issue of area efficiency due to their need to modify the bitcell and integrate the MAC operation into the storage unit, preventing the IMC designs from upscaling to large macros for practical tasks. The modified bitcell requires much more metal layers in the IMC macros, leaving little freedom for global system routing. The system area penalty due to IMC macros will be significant, resulting in a low storage density of those solutions. A huge computational unit is required to handle the AI workloads, which consumes a large area and increases power consumption.  

Fig. \ref{Conv_bittree} shows a block diagram of IMC-based architecture where multiply is performed and later, an adder tree is used to perform the accumulation. The multiplication unit performs the multiplication of inputs and weights. Inputs are applied from outside, and weights are stored in the memory array. As shown in Fig. \ref{Conv_bittree}, the output data is transferred to the adder tree placed outside the macro once the multiplication is performed. For 8-bit precision, one row requires eight routing tracks until the bit tree adder is placed outside the macro. If bit precision increases, the routing connection also increases. The use of XNOR-based multiplication is widely used in SRAM-based in-memory computing cells due to SRAM's fast access times and compatibility with standard CMOS processes \cite{8TXNOR} \cite{XNOR-SRAM} \cite{AICSP}. An 8T SRAM-based digital IMC is used to perform the bit-parallel MAC operation \cite{Mori}. The work proposed an 8-transistor 2b OAI (or and invert) cell that achieves a smaller combined bit cell and multiplier area. However,  designing the OAI circuit itself increases the area overhead. Another work proposed a digital IMC (DIMC) macro using a one-read and one-write 12T bitcell \cite{Fujiwara}. The DIMC macro can realize simultaneous MAC + write
operations and wide-range dynamic voltage-frequency scaling (DVFS) due to the 12T
cell’s 1R1W functionality and low-voltage operation. However, the design of SRAM macro with 12T gives a huge area overhead. X-SRAM \cite{agrawal2018x} proposed an 8T SRAM-based computation, while in our previous work \cite{AICSP}, we performed 10T SRAM-based IMC. This requires an adder tree outside the SRAM macro, which was realized using 28T CMOS full adder, which consumes a huge area overhead and increases the accumulation latency.

To address these issues of general-purpose computing and IMC-based computing in this paper, we propose a Custom Compute Engine-based Architecture that provides improved area efficiency ($TOPS/mm^2$) as compared to in-memory computing architecture and general-purpose computing architectures. This paper extends our work using a proposed 10T SRAM cell to perform the XNOR-based multiply for BNN operation \cite{AICSP}. An adder tree was placed outside the macro, which requires parallel data transfer from the multiplier to the adder tree. This approach requires a massive amount of routing, which consumes a lot of area and introduces latency. Also, massive routing makes it difficult to scale the architecture for bigger array sizes. To address the issues caused by routing, in this work, we propose a custom leaf cell (minimum repetitive block) to design the architecture that overcomes latency, routing efforts and area efficiency.

% \begin{figure}[t]
%     \centering
%     \includegraphics[width=\linewidth]{Figure/routing.png}
%     \caption{Issue of routing in conventional in-memory computing architectures}
%     \label{routing}
% \end{figure}

The major contributions of the work are as follows:
\begin{itemize}
    \item A full adder has been shared with two consecutive cells within the memory array.
    \item The full adder is designed using the minimum cells for low area as compared to conventional CMOS design.
    \item XNOR-based multiplication approach is used for better latency. 
    \item The proposed solution shows improved area efficiency.
\end{itemize}

\section{Proposed In-Memory Computing Architecture}
\begin{figure}[t]
    \centering
    \includegraphics[width=\linewidth]{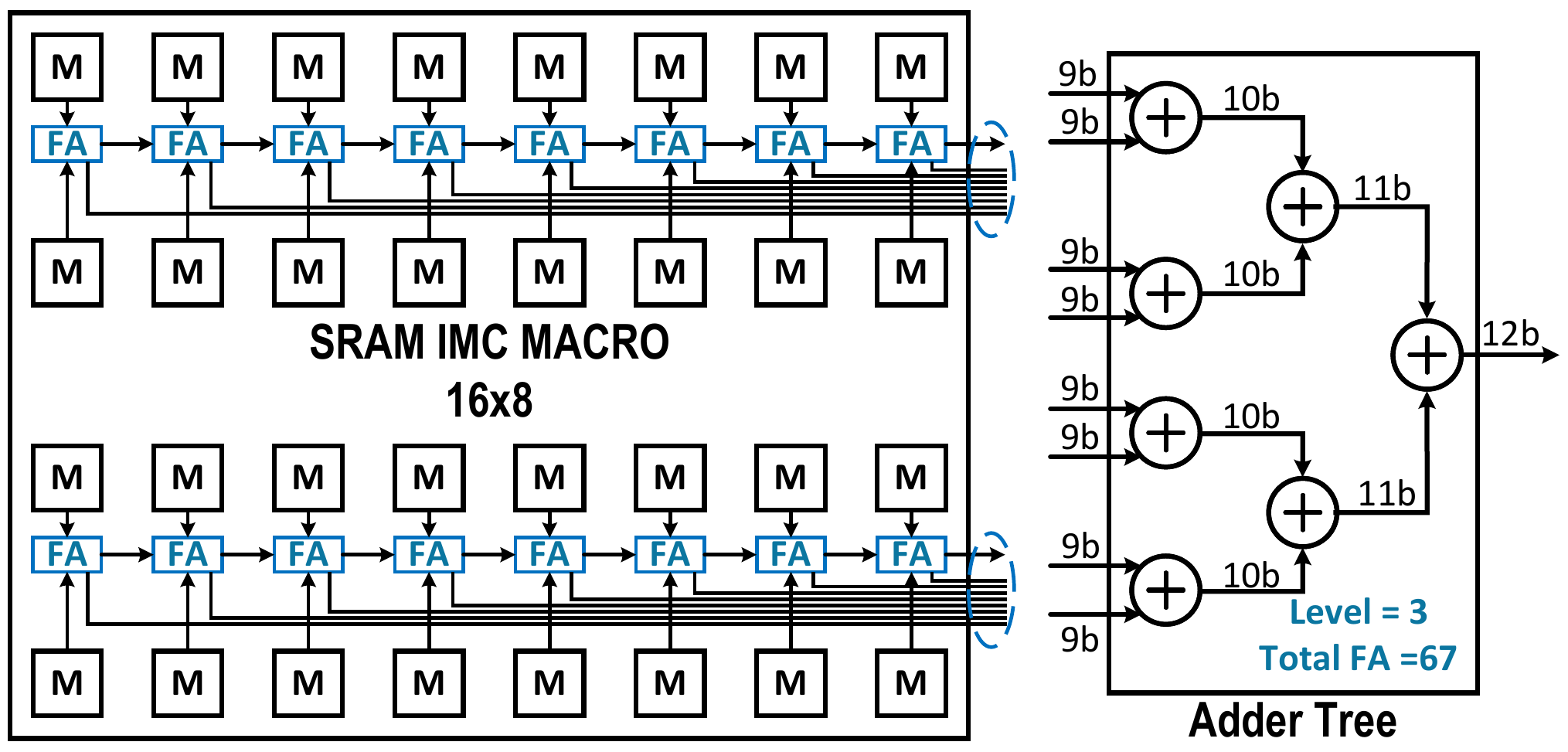}
    \caption{Proposed IMC architecture}
    \label{proposed_bittree}
\end{figure}

Fig. \ref{proposed_bittree} shows the proposed architecture, combining a full adder with two cells of consecutive rows. The multiplication output of two rows is fed as inputs to the full adder. The 8-bit sum output is taken out to the adder tree, and carry is propagated through full adders in the column to get the final carry at the last bit. Finally, we will get a 9-bit output from two rows of SRAM. Bringing the full adder inside the memory macro reduces the latency of accumulation at the first level. It also reduces the routing conjunction, as initially, it requires nine metal routes for two rows to carry the computed MAC to the adder tree (previously, it required 16 routes for two rows). In that case, for an array size of 16$\times$ 8, we require 128 metal tracks, as shown in Fig. \ref{Conv_bittree}.  With the proposed approach, we require only 72 routing tracks to transfer the MAC data from the array to the adder tree, which reduces the latency and area of the IMC macro. In addition to routing track reduction, the proposed approach also reduces one level of the adder tree. Here, we require only 3 level adder tree. Which further reduces the latency and area of the adder tree. The proposed architecture has three major blocks: 1) multiplication unit, 2) full adder inside the macro, and 3) adder tree. Each block is discussed in detail in the later sections. 

\subsection{Read Decoupled 10T SRAM cell}

\subsubsection{Circuit Description}

\begin{figure}[t]
		\centering
		\includegraphics[scale=0.6]{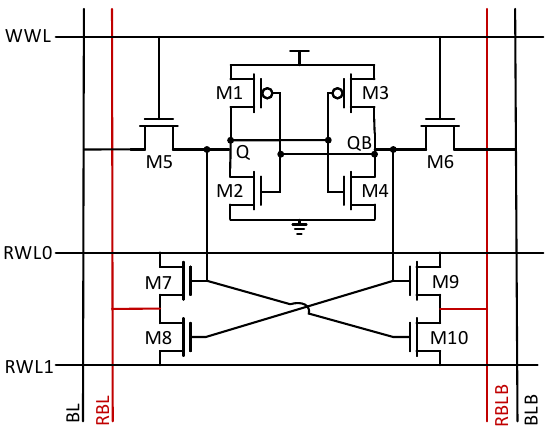}
		\caption{Schematic of the read decoupled 10T SRAM cell}
		\label{proposed_cell}
\end{figure}

Figure \ref{proposed_cell} illustrates the circuit diagram of the read-decoupled 10T SRAM bitcell \cite{AICSP}. The core of the cell consists of a traditional 6T SRAM design for write operations, with four additional transistors (M7-M10) facilitating the decoupled read operation. The gate terminals of M7-M10 are connected to the storage nodes Q and QB and are controlled by two read wordline signals, RWL and RWLB, which interface with the read bitlines RBL and RBLB. This design allows the read and write operations to be performed independently, enhancing the read stability of the cell compared to conventional 6T SRAM cells. 

\subsubsection{Working Principle}
Table \ref{bias_mem} details the control signals for the 10T SRAM cell under various operating conditions. The write operation follows the same procedure as in a conventional 6T cell, where data is written into the cell based on the input applied to BL and BLB, using write access transistors driven by WWL. The hold operation occurs when the cell is not selected, which is determined by detecting no voltage difference on any output bitline. To maintain the cell in a specific logic state during hold, both read bitlines (RBL and RBLB) are precharged to a logic high, and the read wordlines are also kept at a logic high state, preventing any voltage change on the read bitlines. Additionally, the WWL signal is deactivated to disconnect the write bitlines from the storage node, ensuring the cell remains in hold mode.

\begin{table}[t]
\caption{\scshape Biasing for Memory Mode Operations} 
\centering
\resizebox{\linewidth}{!}{
\begin{tabular}{lccccccc}
\toprule                                                                        & \multicolumn{1}{c}{\textbf{WWL}}  & \multicolumn{1}{c}{\textbf{RWL}}         & \multicolumn{1}{c}{\textbf{RWLB}} & \multicolumn{1}{c}{\textbf{BL}}          & \multicolumn{1}{c}{\textbf{BLB}}   & \multicolumn{1}{c}{\textbf{RBL}}      
& \multicolumn{1}{c}{\textbf{RBLB}}  \\ \midrule
\textbf{Hold}   &  L &   H &  H &   X   &    X   &    X &    X  \\
\textbf{Write 0/1}   &    H &    X &    H &    L/H &    H/L &    X &    X  \\
  &    L &    H &    L &    X   &    X   & \multicolumn{2}{l}{   Initialized to H}  \\
\multirow{-2}{*}{\textbf{\begin{tabular}[l]{@{}l@{}}Read 0/1 (RBL)\\ Read 1/0 (RBLB)\end{tabular}}} &    L &    L &    H &    X   &    X   & \multicolumn{2}{l}{Initialized to H}    \\
\bottomrule
\end{tabular}}\\
\vspace{0.5em}
\footnotesize{\scriptsize{H = High, L = Low, X = Don't care}}
\label{bias_mem}
\end{table}

\begin{figure}[t]
		\centering
		\subfigure[Read 0]{\includegraphics[width=0.45\linewidth]{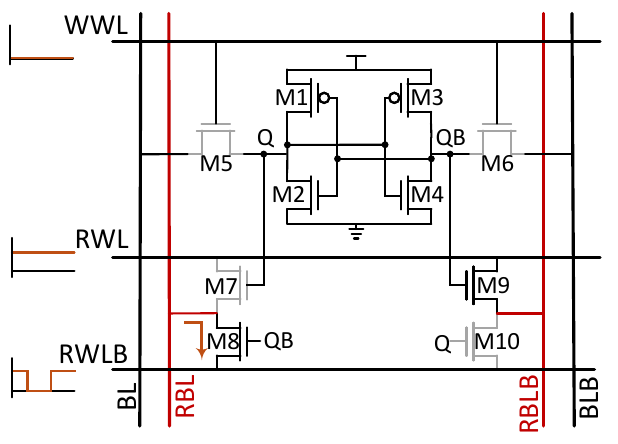}}
		\subfigure[Read 1]{\includegraphics[width=0.45\linewidth]{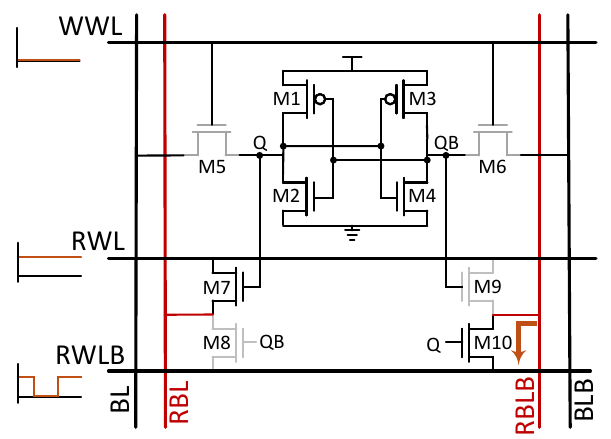}}
		\caption{Read operations for decoupled 10T SRAM cell}
		\label{read_operation}
\end{figure}

To perform a read operation, the read wordline signals RWL and RWLB are used to get the output on the read bitlines, RBL and RBLB. The operating conditions outlined in Table \ref{bias_mem} describe the process of reading data from the proposed cell, covering two scenarios: reading Q on bitline RBL and reading QB on bitline RBLB. Initially, both read bitlines are precharged to a logic high level. The specific inputs applied to RWL and RWLB then determine the output on RBL and RBLB. Figure \ref{read_operation} depicts four cases that illustrate the read-data operation based on the input biasing conditions on the wordlines, RWL and RWLB.

\begin{itemize}
\item Figure \ref{read_operation}(a) illustrates a read-0 operation on RBL, where RWL is set to a logic high level and RWLB to a logic low level. In this scenario, assume $Q=0$ and $QB=1$ are stored in the cell. This configuration turns on transistors M8 and M9 while turning off M7 and M10. Since RBL and RBLB are initially precharged to a high logic level, a conducting path is established between RBL and RWLB via transistor M8, which discharges the voltage on RBL to a low level, resulting in a read-0 (read-Q) operation on RBL with no voltage change detected on RBLB (read-QB operation).

\item Figure \ref{read_operation}(b) depicts a read-1 operation on RBL, with RWL and RWLB set to the same voltage levels as in the previous case. In this instance, assume $Q=1$ and $QB=0$ are stored in the cell. This configuration activates transistors M7 and M10 while turning off M8 and M9. As RBL and RBLB are initially precharged to high logic levels, a conducting path is formed between RBLB and RWLB via transistor M10, discharging RBLB to a low level. This results in a read-0 (read-Q) operation on RBLB, with no voltage change observed on RBL (read-Q operation).
\end{itemize}

If the biasing of wordlines RWL and RWLB is reversed, the read operation functions similarly, but the opposite output is sensed on the bitlines.

% Figure \ref{SNM_compare}(a) compares the static noise margin (SNM) of the proposed 10T cell with that of a conventional 6T SRAM cell. The proposed 10T cell achieves a read SNM of 390mV at VDD = 1V, while the conventional 6T SRAM cell shows a read SNM of 159mV at VDD = 1V. The read SNM (RSNM) of the decoupled 10T cell is 2.45$\times$ of the conventional 6T cell. Figure \ref{SNM_compare}(b) shows the RSNM of the proposed cell under process variation mismatch using Monte Carlo simulation. The read/write latency of the proposed cell was compared with that of a conventional 6T cell at 65nm CMOS technology. Figure \ref{delay} compares the read/write access time across different process corners.

% \begin{figure}[t]
%         \centering
% 		\subfigure[Read access time]{\includegraphics[scale=0.16]{Figure/read delay.png}}
% 		\subfigure[Write access time]{\includegraphics[scale=0.6]{Figure/write delay.png}}
% 		\caption{Latency comparison for read/write operations of 6T SRAM and 10T SRAM cell}
% 		\label{delay}
% \end{figure}

% \begin{figure}[t]
% 	\centering
% 	% \includegraphics[scale=0.13]{Figure/readSNMcells (3).pdf}
%     \subfigure[RSNM comparison]{\includegraphics[scale=0.15]{Figure/readSNMcells (3).pdf}}
% 	% \subfigure[Monte Carlo analysis for RSNM of 10T SRAM cell]{\includegraphics[scale=0.23]{Figure/SNM_MC.png}}
% 	\caption{Read static noise margin analysis}
% 	\label{SNM_compare}
% \end{figure}

\section{In-Memory Binary Convolution} \label{convolution}
% A fundamental aspect of how neural networks operate can be explained using the concept of a dot product. Consider a single unit in a neural network, as illustrated in Figure \ref{neuron}. This unit receives inputs ${X_1, X_2, \dots, X_n}$ from other units and generates an output Y. To compute this output, each input is multiplied by a corresponding weight, represented as ${W_1, W_2, \dots, W_n}$. These weights determine the strength of the connection from each input. The weighted inputs are then summed together, and a bias term, b, is added to the total. The resulting sum is passed through an activation function, F, which defines how the unit responds to the input.

% \begin{center}
    
% $Y = F (\Sigma   (X_i * W_i) + b)$

% \end{center}

% This weighted sum is equivalent to writing like a dot product.
% By the definition of dot product:

% \begin{center}
 
% $\Sigma  X_i * W_i   =   W.X$
   
% \end{center}

% \begin{figure}[t]
% 	\centering
% 	\includegraphics[scale=0.25]{Figure/dp.pdf}
% 	\caption{Multiply and accumulate (MAC) operation}
% 	\label{neuron}
% \end{figure}

\subsection{In-Memory Dot-Product (Multiplication) Operation using XNOR}
Binary Neural Networks (BNNs) were introduced to address this issue by restricting weights and input activations (IAs) to +1 and -1, significantly reducing the need for storage and computation. A more straightforward XNOR-popcount operation replaces the complex dot-product operation in BNNs. When binary values are used for computations, the dot product operation between weights and activation functions can be simplified to bitwise operations, with binary values represented as -1 or +1. These values are encoded using logic '1' for +1 and logic '0' for -1. Table \ref{XNOR multiply} demonstrates how the multiplication of binary values can be interpreted as performing an XNOR operation on these binary-encoded logic values.

\begin{table}[t]
\caption{XNOR Operation Equivalent to Dot-Product}
\centering
\begin{tabular}{ccc}
\toprule
\multicolumn{2}{c}{\textbf{Encoding (Value)}} & \textbf{XNOR (Multiply)} \\ \midrule
 0 (-1) & 0 (-1)    & 1 (+1) \\
0 (-1)  & 1 (+1)    & 0 (-1) \\
1 (+1)  & 0 (-1)    & 0 (-1) \\
1 (+1)  & 1 (+1)    & 1 (+1)  \\ \bottomrule
\end{tabular}
\label{XNOR multiply}
\end{table}

\subsection{In-Memory XOR/XNOR using Read Decoupled 10T SRAM Cell}
\begin{figure}[t]
    \centering
    \includegraphics[scale=0.6]{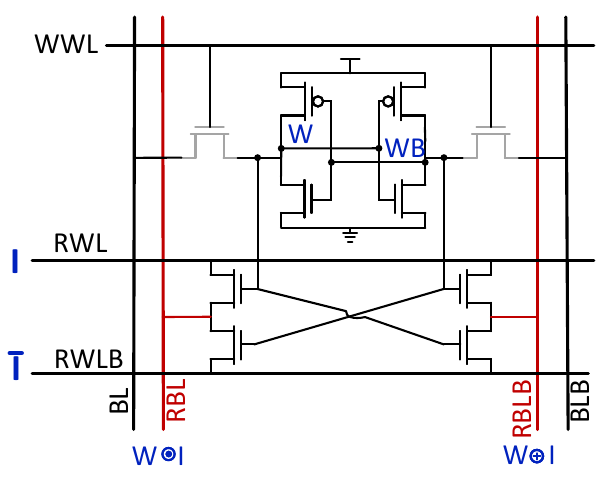}
    \caption{XNOR based Multiplication using 10T SRAM Cell}
    \label{XNOR_MULT_10T}
\end{figure}

The proposed 10T SRAM cell can perform XNOR (multiplication) of the input and weight. The weight (W) is stored in the bitcell and input (I/$\overline{\mbox{I}}$) is applied through RWL/RWLB. Based on stored weights inside memory, transistors discharge to give the required XOR/XNOR output over bitlines (RBL, RBLB). Fig. \ref{XNOR_MULT_10T} shows the schematic of the 10T SRAM cell to perform the XNOR computation and latency comparison with additional XNOR multiplication with 6T and 10T SRAM cells. 

\subsection{Area efficient 14T full adder with full swing for Ripple Carry Adder for Adder Tree}
\begin{figure}[t]
    \centering
    \includegraphics[scale=0.6]{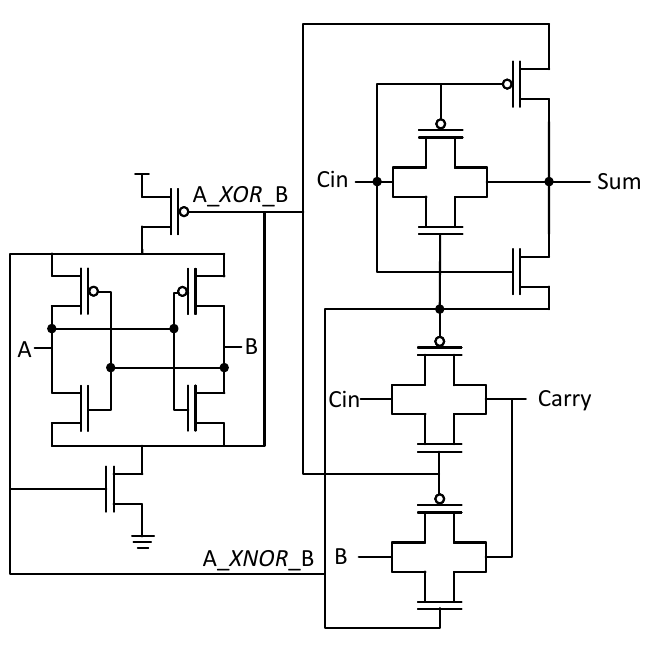}
    \caption[Schematic of 14T full adder]{Schematic of 14T full adder \cite{FullSwing14T}}
    \label{FA14T}
\end{figure}

The adder tree requires a series of full adders to accumulate the parallel output received from the SRAM macro, and it consumes a huge area. Also, it has several levels according to the array size, and the increased number of levels in the adder tree increases the latency. Also, for multi-bit precision, the adder tree requires a multi-bit ripple carry adder at the first level, increasing the latency. As shown in Fig. \ref{Conv_bittree}, the bit-tree adder requires 8-bit adders at the first level, 9-bit at the second level, 10-bit adders at the third level, and 11-bit adders at the fourth level. After that, it gives a 12-bit accumulative output. For this adder tree, if the delay of a 1-bit dull adder is $\delta$, then the overall latency of the bit-tree adder is $4\delta$. Also, it consumes a huge amount of space if designed using a conventional 28T based CMOS design. To overcome the issue of latency, we have performed the first level of accumulation inside the SRAM macro, which reduces the delay of the bit tree adder to $3\delta$ as shown in Fig. \ref{proposed_bittree}. We have taken the minimum 14 transistors-based full adder to reduce the overhead area, providing full swing. Fig. \ref{FA14T} shows the full adder schematic used in this work \cite{FullSwing14T}. The same full adder is used in SRAM macro in between two multiplication unit shown in Fig. \ref{proposed_bittree}.

\section{Results and Performance Comparison}
\begin{figure}[t]
    \centering
    \includegraphics[scale=0.25]{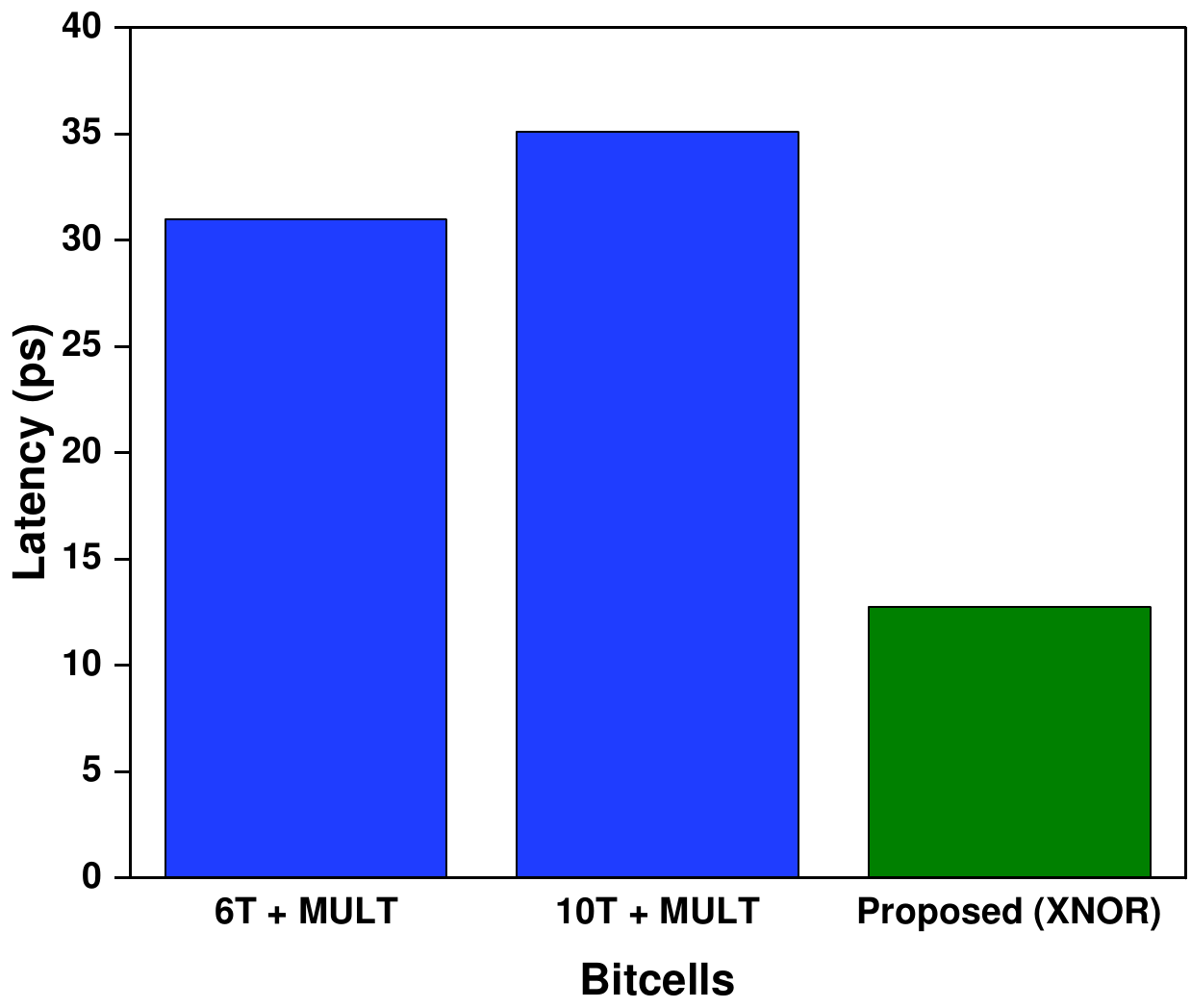}
%     \begin{tikzpicture}[t]
% \begin{axis}[
%      width=6cm,
%      height=6cm,
%     ybar,
%     bar width=10pt,
%     enlargelimits=0.2,
%     legend style={at={(0.5,-0.2)},
%     anchor=north,legend columns=-1},
%     ylabel={Latency (ps)},
%     symbolic x coords={6T+MULT, 10T+MULT, Proposed (XNOR)},
%     xtick=data,
%     nodes near coords,
%     nodes near coords align={vertical},
%     bar width=1cm,]
%     \addplot coordinates {
%         (6T+MULT,30.5) (10T+MULT,35.4) (Proposed (XNOR),12.3)};
% \end{axis}
% \end{tikzpicture}
    \caption{Latency comparison of XNOR based multiplication for different bitcells \cite{AICSP}}
    \label{XNOR_MULT_Latency}
\end{figure}

\begin{figure}[t]
    \centering
    \subfigure[For 1-bit full adder]{\includegraphics[scale=0.25]{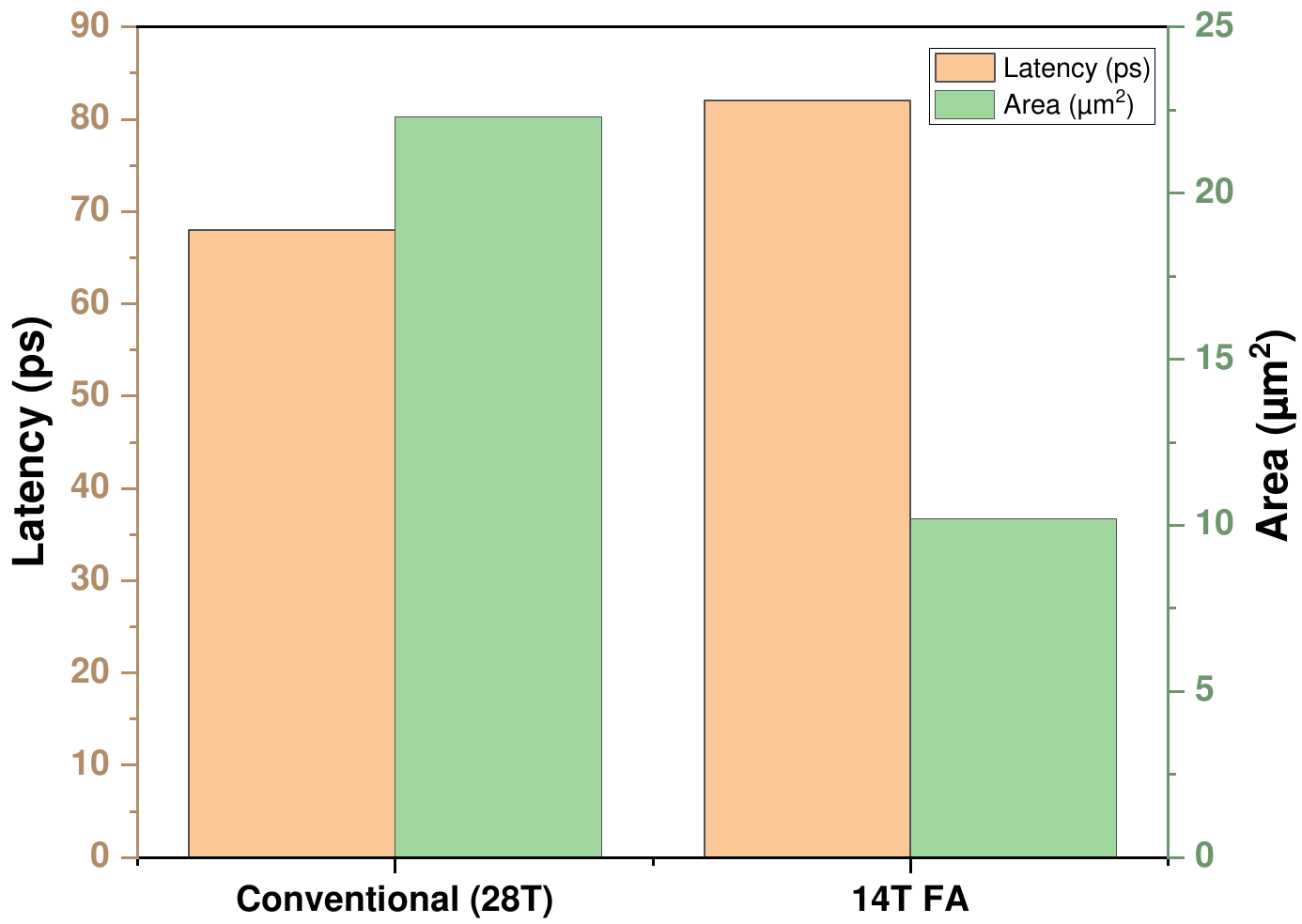}}
    \subfigure[For adder tree]{\includegraphics[scale=0.25]{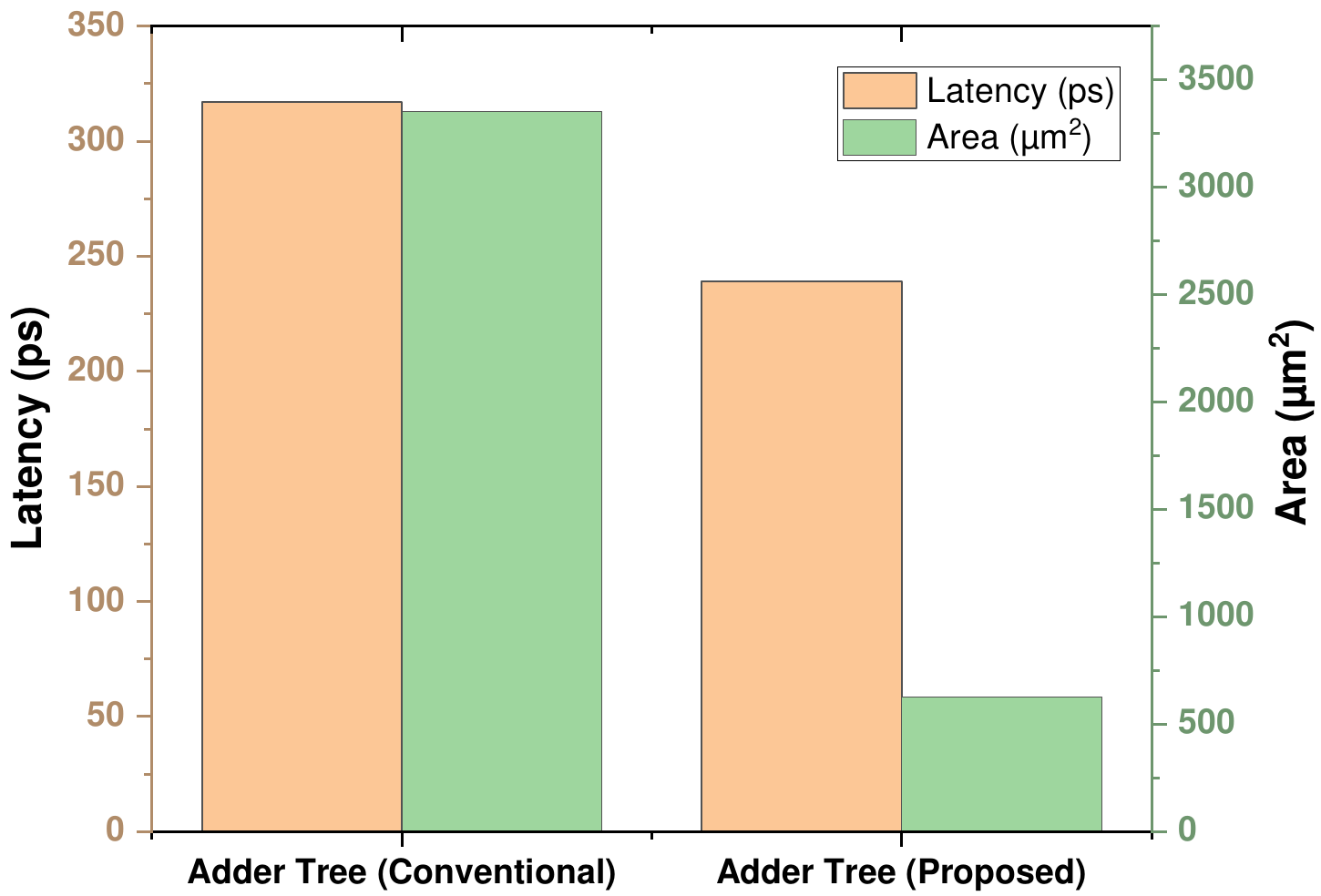}}
    \caption{Latency and area comparisons}
    \label{area_latency_tree}
\end{figure}

\begin{figure}[t]
    \centering
    \includegraphics[scale=0.2]{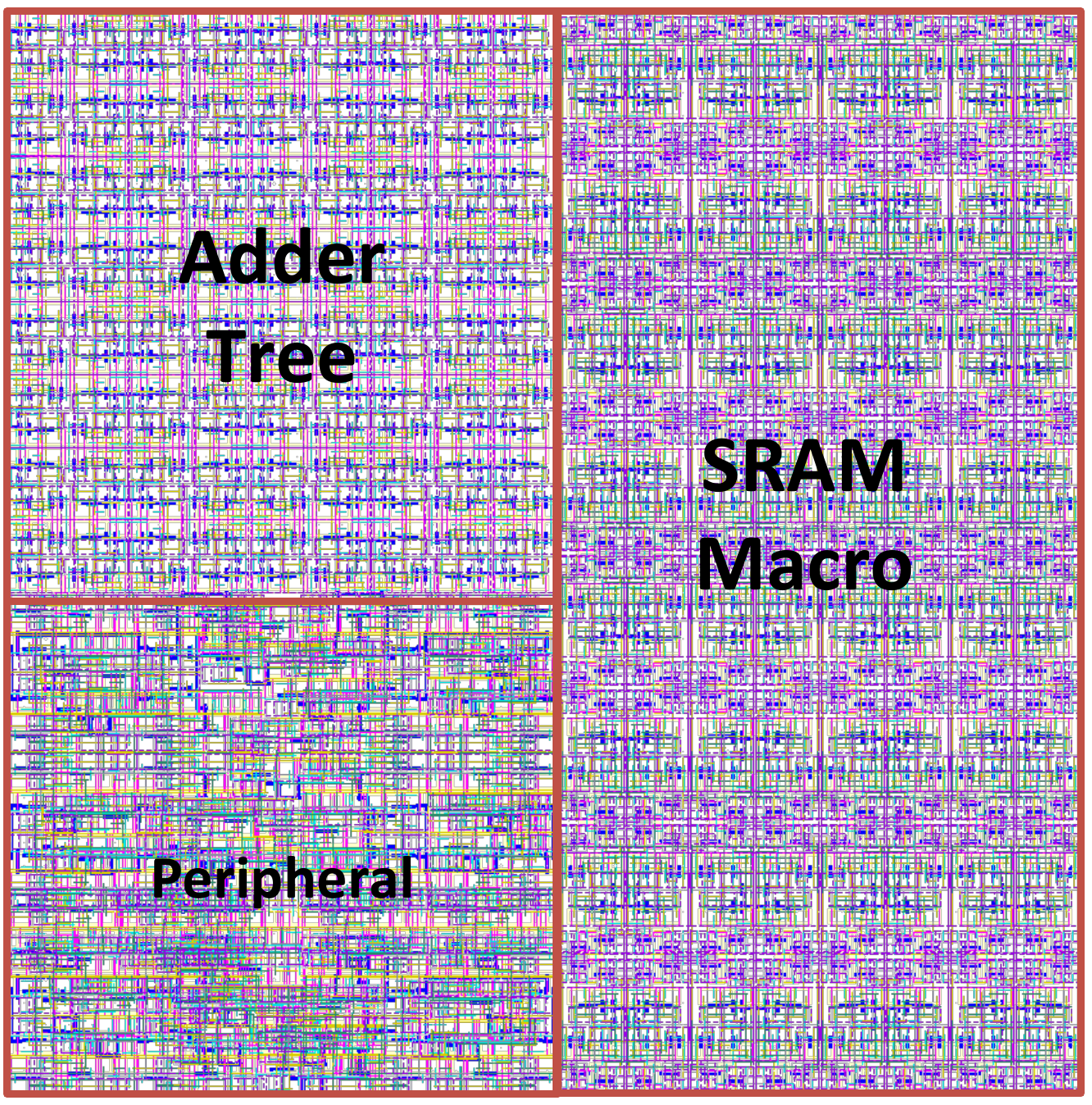}
    \caption{Layout of proposed IMC macro}
    \label{layout}
\end{figure}

\begin{figure}[t]
    \centering
    \includegraphics[scale=0.25]{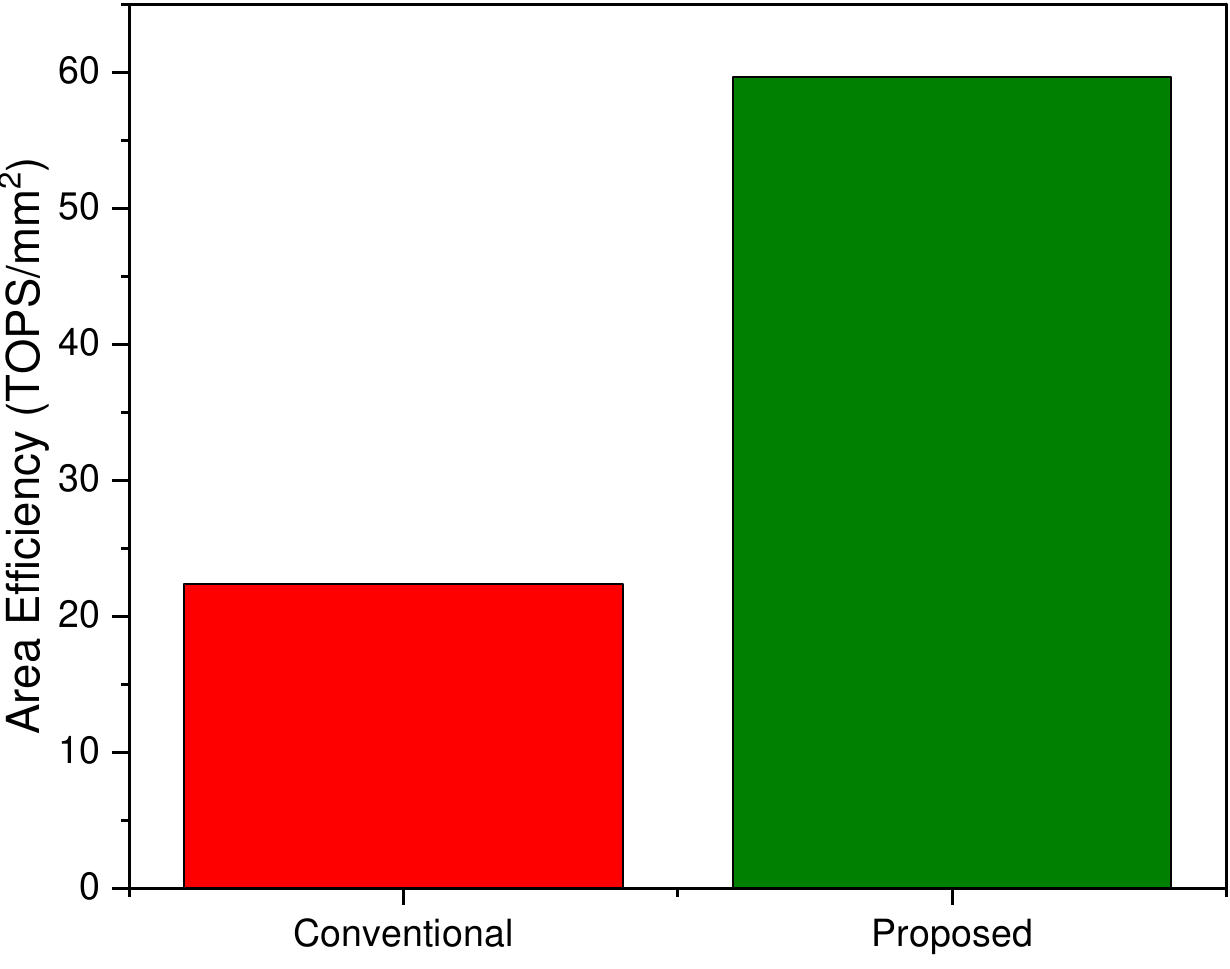}
    \caption{Area efficiency comparison}
    \label{area_efficiency}
\end{figure}

The proposed architecture uses a 65nm CMOS technology node, and post-layout simulations were performed to evaluate the various performance parameters. The Cadence Virtuoso is used to design the schematic and layout, while Calibre is used to extract the parasitic. The performance has been compared for latency, area, and area efficiency. The latency has been compared for the multiplication unit, where we have used 10T SRAM to perform XNOR-based multiplication and compared the latency with conventional 6T/10 SRAM-based XNOR multiplication as shown in Fig. \ref{XNOR_MULT_Latency}. The latency of our design was found to be 58.85\% less as compared to 6T based XNOR. We also compared the latency for the full adder designs using the conventional 28T CMOS design and the 14T full adder design used in this work. The latency of the full adder design used in the architecture is increased by 19\%. However, the area is reduced by 54\% as compared to the conventional CMOS full adder, as shown in Fig. \ref{area_latency_tree}(a). Further, we compared the latency and area of the adder tree as shown in Fig. \ref{area_latency_tree}(b). The results show that the proposed architecture's adder tree latency was reduced by 25\% while the area was reduced by 76\%, which plays a significant contribution to the area efficiency of the macro. Fig. \ref{layout} shows the layout of the proposed IMC architecture. At last, we compared the area efficiency of 16$\times$8 macro for the architectures shown in Fig. \ref{Conv_bittree} and \ref{proposed_bittree}. The area efficiency comparison of the SRAM macro is shown in Fig. \ref{area_efficiency}, which shows 2.67$\times$ improvement. Table \ref{compareTable} shows the performance comparison of the proposed design with the state-of-the-art. The state-of-the-art work has been done using different technology nodes and different array sizes. For a fair comparison, we are comparing area efficiency ($TOPS/mm^2$), which gives the normalized performance.

\begin{table*}[t]
\caption{Performace comparison with state-of-the-art} \label{compareTable}
\centering
\resizebox{\linewidth}{!}{
\begin{tabular}{|l|c|c|c|c|c|c|} \hline
\rowcolor{lightgray}\textbf{Parameter}         & \textbf{\cite{chih}} & \textbf{\cite{Fujiwara}} & \textbf{\cite{Mori}} & \textbf{\cite{multiplyless}} & \textbf{\cite{AICSP}} & \textbf{Proposed Work} \\ \hline
Bitcell  & 6T  & 12T   & 8T   & 8T  & 10T  & 10T \\ \hline
Technology (nm) & 22  & 5  & 4 & 28  & 65  & 65   \\ \hline
Supply Voltage (V)  & 0.72 & 0.5 & 0.9  & 0.9  & 1 & 1 \\ \hline
Operation & MULT+ADD  & MULT+ADD & OAI  & ABS+ADD  & XNOR & XNOR  \\ \hline
Array Size  & 64kb & 64kb  & 54kb & 16kb & 1KB & 16x8 \\ \hline
Precision (input/weight)   & 1/4   & 4/4  & 8/8 & 8/8 & 1/1 & 8/8 \\ \hline
Area Efficiency ($TOPS/mm^2$) & 24.7  & 13.8  & 49.9 & 4.4 & 22.3 & 59.58\\\hline
\end{tabular}}
\end{table*}

\section{Conclusion}
In this paper, we proposed a digital IMC-based compute engine using 10T SRAM-based XNOR computation. The architecture was proposed to reduce the routing conjunction while bringing the parallel multiplication output from macro to the outside adder tree. For that, the architecture brings the adder inside the memory array, which reduces the routing conjunction by half and reduces the adder tree's first accumulation effort. Hence, it reduces the latency and area of the SRAM array and adder tree. Also, we designed the adder tree using a small 14T based full adder, which further significantly reduces the area of the adder tree. The overall architecture shows the 2.67$\times$ area efficiency improvement compared to the state-of-the-art. The significant improvement in area efficiency improves the performance of edge AI hardware.

% \section*{Acknowledgments}
% The authors thank the Ministry of Electronics and Information Technology, Government of India, for providing necessary EDA Tools under the Chips to Startup (C2S) Programme.  

\bibliographystyle{IEEEtran}
\bibliography{Reference}

\end{document}